\documentclass[a4paper, leqno]
{amsart}
\usepackage{natbib}
\usepackage{txfonts}
\usepackage{bbm}
\usepackage{amsmath,amssymb,amsthm,euro} 
\usepackage{bbm}
\usepackage{stmaryrd}
 \usepackage{tensor}
\usepackage{enumerate}
\usepackage{setspace}
\newcommand{\be}{\begin{equation}}
\newcommand{\ee}{\end{equation}}
\newcommand{\bd}{\begin{displaymath}}
\newcommand{\ed}{\end{displaymath}}
\renewcommand{\r}{\mathbb{R}}
\newcommand{\K}{\mathcal{K}}

\newtheorem{theorem}{Theorem}

\newtheorem{corollary}[theorem]{Corollary}

\newtheorem{lemma}[theorem]{Lemma}

\newtheorem{remark}[theorem]{Remark}

\newcommand{\fourIdx}[5]{%
\setbox1=\hbox{\ensuremath{^{#1}}}%
 \setbox2=\hbox{\ensuremath{_{#2}}}%
 \setbox5=\hbox{\ensuremath{#5}}%
 \hspace{\ifnum\wd1>\wd2\wd1\else\wd2\fi}%
 \ensuremath{\copy5^{\hspace{-\wd1}\hspace{-\wd5}#1\hspace{\wd5}#3}%
 _{\hspace{-\wd2}\hspace{-\wd5}#2\hspace{\wd5}#4}%
 }}

\numberwithin{equation}{section}
\numberwithin{theorem}{section}

\begin{document}

\title{  DO ARBITRAGE-FREE PRICES COME FROM UTILITY MAXIMIZATION?   }

\author{Pietro Siorpaes$^{\ddagger}$}
\date{ \today}

  \thanks{We thank Dmitry Kramkov and anonymous referees for their valuable
 comments.}

\thanks{${}^{\ddagger}$ University of Vienna, Faculty of Mathematics,  \emph{email:} pietro.siorpaes@univie.ac.at}

\maketitle

\begin{abstract}
 In this paper we ask whether, given a stock market and an illiquid derivative,  there exists arbitrage-free prices at which an utility-maximizing agent would always want to buy the derivative, irrespectively of his own initial endowment of derivatives and cash.
We prove that this is false for any given investor if one considers \emph{all} initial endowments with finite utility, and that it can instead be true if one restricts to the endowments in the interior.
  We show however how the endowments on the boundary can give rise to very odd phenomena; for example, an investor with such an  endowment would choose not to trade in the derivative even at prices arbitrarily close to some arbitrage price.

\bigskip

\noindent\emph{Keywords:} utility-based pricing, arbitrage free, convex duality, incomplete markets.
\\

\end{abstract}

     \section{Introduction}
    The problem of valuation of a non-traded contingent claim has always been of central importance in mathematical finance. 
One can distinguish between two fundamentally different notions of price: one inherent to the market, and one dependent on the specific investor.
  If a security was sold at an `unreasonable' price,  any investor could lock in a risk-less profit. Even worse, as long as an investor could be considered infinitesimally small with respect to the size of the market, he could create a `money pump' and make this risk-less profit arbitrarily big. These considerations lead to the fecund concept of \emph{arbitrage-free price}, which prescribes a necessary condition for a market model to be viable. 
 
   Fruitful as it was, this notion is of little use to an investor when the mathematical idealization of a complete market falls short of accurately describing reality. Given a non-replicable security, the market mechanism is not sufficient to determine an  interval $I$  of `threshold' prices such that \emph{any} agent should buy at a price smaller than every $p \in I$, sell at a price greater than every $p\in I$, and \emph{do nothing at any price} $p\in I$. Indeed, since the buying or selling at any arbitrage-free price could lead both to a loss or to a gain, the attitude of the agent towards risk must be taken into consideration to decide what he should do at any such price. Intuitively, the interval of threshold prices should depend on the investor and his initial wealth, and it should be contained in the interval of arbitrage-free prices.

   The classical approach in mathematical finance is to assume that the preferences of the agent are determined by the maximal expected utility $u(x,q)$ that he can obtain by investing in the market an initial capital $x$ if holding an endowment consisting of $q$ illiquid contingent claims.
   Pricing rules derived from $u(x,q)$ are called utility-based, and they form the prevailing paradigm in the valuation of contingent claims: see for example  \citet*{Da:97},  \citet*{Frit:00},  \citet*{Fold:00},  \citet*{HodgNeuber:89},  \citet*{KaratKou:96}, \citet*{Kall:02}, \citet*{Hob05a},  \citet*{HeHo09}, \citet*{HugonKramSch:05},  \citet*{KramSirb:06b}.
   
   We will consider the notion of \emph{marginal (utility-based) price} at $(x,q)$, which is defined as a price $p$ such that the (utility-maximizing)  agent with initial endowment $(x,q)$ if given the opportunity to trade the contingent claim at price $p$  would neither buy nor sell any. So marginal prices are precisely the `threshold'  prices previously mentioned; moreover, they are defined also  when considering multiple contingent claims at once.
 The idea underlying this valuation principle is well known in economics: see for example  \citet*{Hick:56}.
  
  This paper is concerned with the investigation of whether all arbitrage-free prices are obtained by utility maximization.
More precisely, we ask if there is a `bad' subset of arbitrage-free prices which are never marginal prices, irrespectively of the investor and his endowment, or if all arbitrage-free prices are equally  `realistic'. And if they are, are they simply parameterized by the initial endowment, or does the preference structure of the investor matter, discriminating between which arbitrage-free prices are among his marginal prices and which arise only because of somebody else? To our knowledge these matters have not been previously investigated.

 What we find is that arbitrage-free prices always come from utility maximization; moreover, it is enough to consider any (one) agent.  
Although  it may sound intuitive that there exists no arbitrage-free price at which every agent should buy (or sell), we recall that, in many occasions, the marginal price at $(x,0)$ is unique and does not depend on $x$ \emph{nor on the utility} function (see  \citet*{KaratKou:96}). Thus, to recover all arbitrage-free prices  it is sometimes not  enough to consider all possible risk aversions:  
we must also take non-zero random endowments. If we do so, it then turns out that we do not need to consider many agents, nor special ones: any (one) utility would do. 
In other words,  we show that there do not exist arbitrage-free prices at which an utility-maximizing agent would always want to buy (or sell) the derivative, irrespectively of his own initial endowment.

However, as we show with an example, there is a very delicate point: one can not a priori discard initial endowments on the boundary of the domain of the utility.
 This will force us to carefully reconsider the framework of  \citet*{HugonKram:04}, on which we rely, and to extend it by including \emph{all} initial endowments with finite utility $u$ (not just the ones in the interior).

    Another objective is to prove that marginal prices are compatible with the market mechanism, i.e., they never allow for an arbitrage. This is important since a negative answer would seriously undermine the legitimacy of this useful concept of price.  \citet*{KaratKou:96} also study this problem, although in a different setting.
   
 We also investigate a number of properties of marginal prices, with special emphasis  on the peculiarities which arise when dealing with endowments on the boundary of the domain of $u$, as these points are essential to answer positively our main question.
   As shown by  \citet*{KaratKou:96} (and, in our framework, by  \citet*{HugonKramSch:05}), under appropriate assumptions, the marginal price at $(x,0)$ is unique; so if an agent has no initial endowment of illiquid contingent claims,  the interval $I$ of marginal prices collapses to a point, a particularly pleasant situation.
 In stark contrast to this, we prove that marginal prices based at a point $(x,q)$   on the boundary, if they exist, are never  unique; we also characterize
  their existence, and show how to compute them.
   Moreover, we give an explicit example of non-uniqueness where the set of marginal prices contains an interval that has an \emph{arbitrage price} as an end-point.
We then prove that this considerably quirky behavior is a general fact; in particular,  given an investor with an  endowment on the boundary, there always is an  arbitrage price $p$ such that the agent  would  choose not  to trade in the derivative even if this one was being sold at a price arbitrarily close to $p$! 
Thus, one is left with a choice: either  to restrict the attention to the well behaved marginal prices  based at endowments in the interior of the domain of $u$ (in which case however  arbitrage-free prices may not come from utility maximization),  
or to consider also marginal prices with awkward properties.

  The plan of the rest of the paper is as follows: 
in Section \ref{themodel} we introduce the model we work in, and
in Section \ref{Statement of the main theorem} we state our main theorems.
Then, in Section   \ref{counterexample} we show with an explicit example what can go
  wrong, and in Section \ref{Characterizations of Arbitrage-Free Prices} we present a useful   characterization of arbitrage-free prices. Finally, in Section \ref{u upper semi-continuous} we obtain that the maximal expected utility  $u(x,q)$ is an upper semi-continuous function,  
  in Section \ref{Proof of the main theorem.} we prove our first main theorem, and  in Section \ref{Consequences and related results} 
 we derive a formula for marginal utility-based prices and we prove our second main theorem.

     \section{The model}
     \label{themodel}
We use  the same model of an agent investing in a financial market as  \citet*{HugonKram:04}, and the same notations; however, we have to consider slightly more general random endowments, or our main result would not hold (see Section \ref{counterexample}). We thus present only a very brief introduction to the model, referring to  \citet*{HugonKram:04} for additional details and discussions, and pointing out the few instances where we differ from it. 
       
    We consider a model of a financial market  composed of a savings account with zero interest rate, and $d$ stocks with price $S = (S^i)_{i=1}^d$. We assume that  $S$ is a  locally-bounded\footnote{This assumption is not really necessary, as the results in \citet*{KramSch:99}, \citet*{KramSch:03}, \citet*{HugonKram:04},  \citet*{DelbSch:97} on which our proofs hinge, although proved for a locally-bounded semi-martingale, are true also without the local boundedness assumption, if one replaces equivalent local-martingale measures with separating measures throughout. This observation is stated in \citet*[Remark 3.4]{HugonKramSch:05}.}
 semi-martingale on a filtered probability space $(\Omega,\mathcal{F},\mathbb{F},P)$ whose filtration $\mathbb{F}=(\mathcal{F}_t)_{t\in [0,T]}$ satisfies the usual conditions. 
  The wealth $ X$  of a (self-financing) portfolio $(x,H)$ evolves in time as the  stochastic integral
\bd
\label{wealth} \textstyle
X_t=x+(H\cdot S)_t=x+\int_0^t H_u dS_u , \qquad t\in [0,T] ,
\ed
where $H$ is assumed to be a predictable $S$-integrable process.
We denote by $\mathcal{X}(x)$ the set of non-negative wealth processes whose
initial value is equal to $x\geq0$, and by $\mathcal{M}$ the 
family of equivalent local martingale measures; we assume that 
\be
\label{na}
\mathcal{M}\ne \emptyset .
\ee
 The  utility function $U$ is assumed to be strictly concave, strictly increasing and continuously differentiable on $(0,\infty)$ and to satisfy Inada conditions:
\be
\label{inada} \textstyle
U'(0+):=\lim_{x \to 0+}U(x)=\infty  , \qquad
U'(\infty):=\lim_{x \to \infty}U'(x)=0 .
\ee
Differently from \citet*{HugonKram:04}, it will be convenient for us to consider $U$ as defined on the whole real line. We want its extension to be concave and upper semi-continuous, and  (\ref{inada}) implies that there is only one possible choice: we define $U(x)$ to be $-\infty$ for $x$ in $(-\infty,0)$, and to equal $U(0+)$ at $x=0$.

In this paper we will be concerned with a family of non-traded European contingent claims whose payoff $f=(f_j)_{j=1}^n$ is  dominated  by the final value of a  non-negative wealth process $X'$, that is,
\be
\label{domination} \textstyle
|f|:=\sqrt{\sum_{j=1}^n |f_j|^2 }\leq X'_T  \, .
\ee
 To rule out doubling strategies in the model, one has to impose some sort of boundedness condition on the allowed wealth processes.  In the presence of contingent claims bounded with respect to some numéraire (i.e., satisfying \eqref{domination}), one should consider wealth processes that are admissible under some numéraire; these can be characterized as follows.
 A process $X$ in $\mathcal{X}(x)$ is said to be \emph{maximal} if its terminal value cannot be dominated by that of any other process in $\mathcal{X}(x)$. 
A wealth process $X$ is called \emph{acceptable} if it admits a representation of the form 
$X = X' - X''$, where $X'$
is a non-negative wealth process and $X''$ is a maximal wealth  process.
 
 Consider an agent with an initial endowment consisting of a cash amount $x$ and a quantity $q$ of contingent claims $f$. If the agent  followed the strategy $H$ his wealth process would be $X=x+H\cdot S$ and his final wealth  $X_T+qf$ (throughout this paper we will use the notation $vw$ for the dot product of the vectors $v$ and $w$).
   If the contingent claims $f$ cannot be traded and all the agent can do is to  invest in the stocks and the bond, then his maximal expected utility will be 
 \be
 \label{prrend}
 u(x,q):=\sup_{ \{X :\, X \textrm{ is acceptable}, \, X_0 = x\} } \mathbb{E}[U(X_T+qf)] ,   \qquad (x,q)\in \r\times \r^n
 \ee
 where we define $\mathbb{E}[U(X_T+qf)]$ to be $-\infty$ when $\mathbb{E}[U^{-}(X_T+qf)]=-\infty$, whether or not $\mathbb{E}[U^{+}(X_T+qf)]$ is finite (so, unlike  \citet*{HugonKram:04}, we define $u$  for every $(x,q)\in \r\times \r^n$). Clearly, if the wealth process does not satisfy $X_T+qf\geq 0$,  the corresponding expected utility will be $-\infty$, so we set
\bd
\mathcal{X}(x,q):=\{ X : \,\, X \textrm{ is acceptable}, X_0 = x \textrm{ and } X_T+ qf  \geq 0 \} .
\ed
Notice that $\mathcal{X}(x,0)=\mathcal{X}(x)$ for all $x\geq 0$. We will consider the convex cone
\be
\label{defK}
\mathcal{\bar{K}}:=\{(x,q)\in \r \times \r^{n}: \mathcal{X}(x,q)\ne \emptyset \} .
\ee 
 which is closed and  contains $(1,0)$ in its interior $\mathcal{K}$  (see  \citet*[Lemma 6 and Lemma 1]{HugonKram:04}). Being convex, $\mathcal{\bar{K}}$ is then the closure of its interior  $\mathcal{K}$.

  We point out that, if conditions (\ref{na}), (\ref{inada}), and (\ref{domination}) hold and 
$ u(x,0)<\infty$ for some $x>0 $,
 then the concave function $u$ defined on $\r^{n+1}$ by (\ref{prrend})
 never takes the value $\infty$ and $\mathcal{K}\subseteq \{u >-\infty
 \} \subseteq \mathcal{\bar{K}}$. In particular $\mathcal{K}$ is the interior of $\{u >-\infty \}$, and $-u$ is a proper convex function.

 Suppose now that it becomes possible to trade the contingent claims \emph{at time $0$} and at price $p$. Consider an  agent with random endowment $(x,q)$ who buys $q'$ contingent claims, spending $q' p:=\sum_{j=1}^n q'_jp_j$. If he then invests the remaining wealth $x-q'p$ dynamically into the stocks and bond following the strategy $H$, his wealth process will be $X=x-q'p+(H\cdot S)$ and his final wealth will be $X_T+(q+q')f$.
 
 Of course, we want to rule out arbitrage possibilities in the expanded market.
We will say that $p$ is an \emph{arbitrage-free price} for the European contingent claims $f$ if any portfolio with
zero initial capital and non-negative final wealth has identically zero final wealth, and we will denote by $\mathcal{P}$ the set of arbitrage-free prices; so, we set
 $$\mathcal{P}:=\{ p \in \r^n :  q\in\r^n, X\in\mathcal{X}(-pq,q) \text{ imply } X_T=-qf\}.$$

 Consider now an agent with utility $U$ and with corresponding maximal expected utility $u$ given by  (\ref{prrend}), and fix  
 a point $(x,q)$ in $\{u >-\infty \}$.
We will say that, $p$ is a \emph{marginal (utility-based) price} at $(x,q)$ for  $f$ relative to $U$  if the agent with initial endowment $(x,q)$ given the opportunity to trade the contingent claims $f$ at time zero at price $p$ would neither buy nor sell any. We will denote by  $\mathcal{P}(x,q; U)$ (or simply $ \mathcal{P}(x,q)$) the set of marginal prices of $f$  at $(x,q)\in \{ u\in \r\}$; i.e., we set
 \be
 \label{mubp} 
 \mathcal{P}(x,q; U):=\{p\in \r^n :  u(x-q'p,q+q') \leq u(x,q)   \text{ for all } q' \in \r^{n}\}.  
\ee
  Intuition suggested that in the previous definition we exclude a priori the points $(x,q)$ in $\{u=-\infty \}$; our main result will show that indeed this is a good choice.
    
\section{Statement of the main theorem}
\label{Statement of the main theorem}
 In this section we state our main results; our first objective is to answer to the following questions:
  \begin{enumerate}
  \item
  Can one span all arbitrage-free prices using marginal utility-based prices? In particular, does one need to consider the marginal utility-based prices at all points $(x,q)$ in $\{ u>-\infty\}$, or is it enough to consider the `nicer' points in the interior? Does one have to take the prices relative to all possible utilities, or to some subfamily, or is it enough to consider any one utility function?
    \item
   It is true that all marginal utility-based prices are arbitrage-free?
  \end{enumerate}
 
We now need to introduce some more (standard) notation. We set
\be
\label{Y}
\mathcal{Y}(y):=\{Y\geq0 : Y_0 = y , XY \textrm{ is a super-martingale for all } X\in \mathcal{X}(1)\}, 
\ee
and we denote with $V$  the convex conjugate of $U$, i.e.,
\be
\label{V} \textstyle
V(y):=\sup_{x \in \r }(U(x)-xy)=\max_{x >0 }(U(x)-xy), \quad y \in \r.
\ee

 Following \citet*{HugonKram:04}, we will denote by $w$  the value function of the problem of optimal investment without the European contingent claims, and by $\tilde{w}$  its dual value function; in other words
 \be
 \label{w}
 w(x):=\sup_{X \in \,\, \mathcal{X}(x)} \mathbb{E}[U(X_T)] , \quad \tilde{w}(y):=\inf_{Y\in\mathcal{Y}(y)}\mathbb{E}[V(Y_T)] . 
\ee
       
Following \citet*{DelbSch:97}, we will say that a wealth process $X$ is \emph{workable} if both $X$ and $-X$ are acceptable, and following \citet*{HugonKramSch:05} we will say that a random variable $g$ is \emph{replicable} if there is an workable process $X$ such that $X_T = g$.
 Provided that it exists, such a process $X$ is unique.
To simplify some proofs we will also  assume that
\be
\label{wlog}
\textrm{ for any non-zero $q\in \r^n$ the random variable $qf$ is not replicable} ;
\ee
we remark however that most of our result are clearly valid without this assumption (for reasons explained in \citet*[Remark 6]{HugonKram:04}).

The following theorem constitutes our first main result.

\begin{theorem}        
\label{afp=mubp}
  Assume that conditions (\ref{na}), (\ref{inada}), (\ref{domination}) and (\ref{wlog})  hold and that
 \bd
 \tilde{w}(y) <\infty \textrm{  for all } y>0, 
 \ed
 where $\tilde{w}$ is the function defined by (\ref{Y}), (\ref{V}), (\ref{w}).
     Then the set of arbitrage-free prices coincides with the set of all marginal utility-based prices relative to the utility function $U$. In other words 
    \be
    \label{eqafp=mubp}
    \qquad  \mathcal{P}=\bigcup_{(x,q)\in \{ u >-\infty \}}  \mathcal{P}(x,q;U) ,
    \ee
    where $u$ is the function defined in (\ref{prrend}).
\end{theorem}
Theorem \ref{afp=mubp} (which we will prove in Section \ref{Proof of the main theorem.}) says in particular that marginal prices are always arbitrage free, and to recover all arbitrage-free prices with marginal prices there is no need to consider multiple utilities, nor any special one: any one utility function will do!

   We will see in Section \ref{counterexample} that, to obtain a result like the above, it is \emph{not} sufficient to consider the points $(x,q)$ in the interior $\mathcal{K}$ of $\{ u >-\infty \}$; this is why we are forced to extend the framework of  \citet*{HugonKram:04}.
This leads us to study marginal prices based at boundary points, which turn out to behave quite unexpectedly. What is perhaps their most quirky  property is described in our second main theorem, where by 
$[p_0,p)$ we denote the segment from $p_0$ to $p$, excluding $p$.

\begin{theorem}  
   \label{Pnotclosed}
   Under the assumptions of Theorem \ref{afp=mubp}, if  $p_0 \in \mathcal{P}(x,q)$ for some non-zero $(x,q)$ on the
   boundary of $ \{ u >-\infty \} $, then there exists an
   \emph{arbitrage price} $p \in \r^n \setminus \mathcal{P}$ such
   that $[p_0,p) \subseteq \mathcal{P}(x,q)$.
\end{theorem}
Note how Theorem \ref{Pnotclosed} (which we will prove in  Section \ref{Consequences and related results}) also stands in stark contrast with the
fact that the marginal price at $(x,0)$ is unique (under appropriate assumptions), as shown by  Hugonnier, Kramkov and Schachermayer \citet*{HugonKramSch:05}. To the best of our knowledge, no assumptions are known under which the marginal prices based at the generic point  $(x,q)$ in the    interior of $ \{ u >-\infty \} $ are unique (or, equivalently, $u$ is differentiable).

 \section{An illuminating example}
 \label{counterexample}
We will now construct explicitly an illuminating example which 
shows why we need all endowments with finite utility to prove Theorem \ref{afp=mubp}, and  illustrates how a number of intuitive statements about marginal prices fail to be true for prices based at  boundary points,  clarifying  in particular why Theorem \ref{Pnotclosed} holds.

  We will denote by $(1,p)^{\perp}$ the set of vectors in $\r^{n+1}$ orthogonal to $(1,p)$.
  Given any $x\in \r$ and the set $\mathcal{\bar{K}}$
  defined in (\ref{defK}), if we define 
     \be
     \label{optmset}
     \mathcal{\bar{K}}^p(x):=\mathcal{\bar{K}} \cap \{ (x-qp,q): q\in \r^n \}, 
     \ee
     then $\mathcal{\bar{K}}^p(x)$ is the set of points $(a,b) \in \mathcal{\bar{K}}$ whose dot product with $(1,p)$ equals $x$,  i.e., the intersection of $\mathcal{\bar{K}}$ and the set $(x,0)+ (1,p)^{\perp}$.  
   Since $u=-\infty$ outside $\mathcal{\bar{K}}$, for $(a,b)\in\{ u\in \r\}$
  it follows from the definition of marginal prices (\ref{mubp}) that $p$ \emph{is a marginal price at} $(a,b)$  \emph{if and only if} $(a,b)$ \emph{is a maximizer of} $u$ \emph{on} $\mathcal{\bar{K}}^p(a+bp)$.
  
 Let us now build an example of the following counter-intuitive
 situation: an arbitrage-free price $p$ and a maximal expected utility function $u$ that, for arbitrary $x\in \r$ s.t.  $\mathcal{\bar{K}}^p(x)\neq \emptyset$, when restricted to  $\mathcal{\bar{K}}^p(x)$ attains its maximum \emph{only} at some point $(x',q')$ on the boundary of $\mathcal{\bar{K}}$. This will give that $p$ belongs to $\mathcal{P}(x',q')$ but not to any $\mathcal{P}(\bar{x},\bar{q})$ with $(\bar{x},\bar{q})$ in $\mathcal{K}$ (since $(\bar{x},\bar{q})\in \mathcal{\bar{K}}^p(\bar{x}+\bar{q}p)$).
 
 Take $U(x)=\sqrt{x}$, and consider a market with the bond, no stocks and one contingent claim whose law has density 
 \bd
 c(s+1)^{\frac{3}{2}} \, 1_{[-1,1]}(s)
 \ed
 with respect to the Lebesgue measure, where $c$ is the normalization
 constant. In this case $\mathcal{\bar{K}}=\{(a,b)\in \r^2 : |b| \leq
 a \}$, and trivially $(-1,1)$ is the set of arbitrage-free prices
 $\mathcal{P}$. Fix $p=0$ and\footnote{The other cases are trivial: if $x<0$ then $\mathcal{\bar{K}}^p(x)=\emptyset$, and if $x=0$ then $\mathcal{\bar{K}}^p(x)=\{0\}$, which is a subset of the boundary of $\K$.}  $x>0$, then  $\mathcal{\bar{K}}^p(x)=\{x\} \times [-x,x]$ and we want to prove that the concave function 
 \bd \textstyle
 h(q):=u(x,q)=c\int_{-1}^1 \sqrt{x+qs}\, \left(s+1\right)^{\frac{3}{2}} ds , \quad q\in [-x,x] ,
 \ed
   does not attain a maximum at any point in the interior $(-x,x)$ of $[-x,x]$. By differentiating under the integral we compute the derivative of $h$ to be 
 \bd 
 h'(q)=c\int_{-1}^1 \frac{s(s+1)^{\frac{3}{2}}}{ 2\sqrt{x+qs}} ds .
 \ed
 The function $h$ is concave, so its derivative is non-increasing. It follows that since
 \bd \textstyle
 h'(x)=\left( c\int_{-1}^1 s(s+1) ds \right) / 2 \sqrt{x}=c/3\sqrt{x}
 \ed  
  is strictly positive, $h$ is strictly increasing on $[-x,x]$, so its maximum is attained only at $q=x$.
 Thus $p=0 \in \mathcal{P}$ belongs to $\mathcal{P}(x,x)$ (for any $x>0$) but not to
 any $\mathcal{P}(\bar{x},\bar{q})$ with $(\bar{x},\bar{q})$ in
 $\mathcal{K}$, proving that it is not enough to consider endowments in the interior of $\{ u>-\infty \}$.

This example also illustrates how the uniqueness of
  marginal utility-based prices fails on the boundary; let us show
that indeed  $(-1,0) \subseteq \mathcal{P}(x,x)$.
 If $-1<r<0$ we have that (draw a picture!)
  \be
\label{nonunique}
   \mathcal{\bar{K}}^r(x(1+r))= \{(a,b)\in \mathcal{\bar{K}}: a+br=x(1+r) \} \subseteq  \{(a,b)\in \mathcal{\bar{K}}: a\leq x \} \, .
  \ee
It is clear\footnote{Indeed
$ u(\cdot,q)$ is obviously increasing, and we have seen above that the maximum of  
$u$ on  $\{x\} \times [-x,x]$ is  attained at the point $(x,x)$.}  that the maximum of $u$ on the set $\{(a,b)\in
  \mathcal{\bar{K}}: a\leq x \}$ is attained at $(x,x)$. Thus  (\ref{nonunique}) implies that the vector $ (x,x)\in  \mathcal{\bar{K}}^r(x(1+r))$  is a maximizer of $u$ on
$\mathcal{\bar{K}}^r(x(1+r))$, i.e., $r\in \mathcal{P}(x,x)$.

In this example not only $\mathcal{P}(x,x)$  is not a singleton: it
even contains a whole open interval of prices with one extreme being
an arbitrage price. This is an example of the general behavior described in Theorem \ref{Pnotclosed}.

The inclusion $(-1,0] \subseteq \mathcal{P}(x,x)$, valid for all $x>0$, also shows that the same marginal price can correspond to multiple initial endowments.
 So marginal prices cannot be thought of as a parametrization---nor as a partition---of arbitrage-free prices, with the parameter being the initial endowment.
  
  Observe how this example yields a function $u$ which is finite and continuous on $\mathcal{\bar{K}}$. While this is not true in  general, we will see that $u$ is always upper semi-continuous.

 \section{Characterizations of Arbitrage-Free Prices}
   \label{Characterizations of Arbitrage-Free Prices}

We will need the following definition and technical lemma
\begin{align}
\label{primalSetC}
\mathcal{C}(x):=\{ g \in L_+^0(P) : g \leq X_T \textrm{ for some } X\in \mathcal{X}(x)\} .
\end{align} 

\begin{lemma}      
\label{existswork}
  Assume condition (\ref{na}). Then given any $X \in \mathcal{X}(x,q)$ there exists a workable $\tilde{X}\in \mathcal{X}(x,q)$ such that $\tilde{X}_T\geq X_T$.
\end{lemma}

  \textsf{ PROOF. }  Since $\mathcal{C}(x)$  is closed and bounded in $L^0(P)$ (see \citet*[Proposition 3.1]{KramSch:99}). Let $X=X'-X''$ be a decomposition of the acceptable process $X\in \mathcal{X}(x,q)$ into a nonnegative wealth process $X'$ and a maximal process $X''$, and define
\bd
B:=\mathcal{C}(x) \cap \{ g \in L^0(P) : g \geq X'_T \} .
\ed
 The set $B$ is then closed and bounded in $L^0(P)$ and so, as stated and proved in the course of the proof of lemma 4.3 in \citet*{DelbSch:94}, $B$ contains a maximal element $\tilde{g}$. It follows that there exists a maximal $\tilde{X}'\in  \mathcal{X}(x)$ such that $\tilde{X}_T^{'}=\tilde{g}\geq X'_T$. Take $\tilde{X}:=\tilde{X}'-X''$, then $\tilde{X}$ is workable and $\tilde{X}_T\geq X_T$, so $\tilde{X}\in \mathcal{X}(x,q)$. $\Box$
 \newline

    We will denote by $\mathcal{\bar{L}}$ the polar of $-\mathcal{\bar{K}}$, i.e.,
\be
\label{L}
\mathcal{\bar{L}}:=-\mathcal{\bar{K}}^o:=\{ v\in\r^{n+1} : vw \geq 0 \textrm{ for all } w \in \mathcal{\bar{K}} \} .
\ee
 Clearly, $\mathcal{\bar{L}}$ is a closed convex cone; we will denote its relative interior by $\mathcal{L}$.
The following facts will be used in the proof of Lemma \ref{xacc0}.
 As shown in \citet*[Lemma 7]{HugonKram:04}, assumptions (\ref{na}),
 (\ref{domination}) and (\ref{wlog})  imply that  $\mathcal{L}$ is the interior of  $\mathcal{\bar{L}}$, and so
\begin{align}
\label{LK>0}
\mathcal{L}=\{(y,r)\in \r^{n+1} : xy+qr>0  \text{ for every non-zero } (x,q)\in \mathcal{\bar{K}} \} 
\end{align} 
and  $\mathcal{\bar{K}} \cap (-\mathcal{\bar{K}})=\{0\}$ (since   \citet*[Corollary 14.6.1]{Rock:70} implies that
 $\mathcal{\bar{K}} $ contains no line passing through the origin).
We recall that the recession cone\footnote{Also known as asymptotic cone.} rec(C) of a set $C\neq \emptyset$ is defined as the set of all $y$ such that $x+ty\in C$ for all $x\in C$ and $t\geq 0$; we refer to \citet*[Theorem 8.4]{Rock:70} for the following fundamental property of recession cones:
\begin{align}
\label{RecConeBdd}
\hspace{-0.2cm}\text{a closed convex set C$\neq \emptyset$ is unbounded iff there exists a non-zero } x\in \text{rec(C)}.
\end{align}

\begin{lemma}  
 \label{xacc0}
Assume conditions (\ref{na}), (\ref{domination}) and (\ref{wlog}). Then the following are equivalent:
  \begin{enumerate}
\item  \label{Pagain} $p$ is an arbitrage-free price.
\item  \label{Lagain} $(1,p)\in \mathcal{L}$.
\item \label{E_Q(f)=p} There exists an equivalent local martingale measure $Q\in \mathcal{M}$ that  satisfies $\mathbb{E}_Q[f]=p$ and such that the maximal process $X'$ that appears in (\ref{domination}) is a uniformly integrable martingale under $Q$. 
\item \label{perp=0} $ \mathcal{\bar{K}} \cap (1,p)^{\perp}=\{0\}$.
\end{enumerate}
Moreover, if $x\in \r$ is such that the set $\mathcal{\bar{K}}^p(x)$ defined in (\ref{optmset}) is non-empty, then its recession cone is  $ \mathcal{\bar{K}} \cap (1,p)^{\perp}$, and so another equivalent condition is
\begin{enumerate}  \setcounter{enumi}{4}
\item \label{Kpbdd}
The set $\mathcal{\bar{K}}^p(x)$ is bounded.
\end{enumerate}

 \end{lemma}
  
  The previous lemma shows that item \ref{E_Q(f)=p}, which would have been a plausible alternative definition of arbitrage-free price (and was implicitly used in \citet*{HugonKram:04}), is actually equivalent to our definition.    \newline
  
  \textsf{ PROOF. }  
Item \ref{Lagain} implies item \ref{perp=0}, as it follows from \eqref{LK>0}.  To prove the opposite implication take $(1,p)\notin \mathcal{L}$, so that \eqref{LK>0} implies that there exists a non-zero $(\tilde{x},\tilde{q})\in \mathcal{\bar{K}}$ such that $\tilde{x}+p\tilde{q} \leq 0$.
We can then find a convex combination $(x',q')$  of $(\tilde{x},\tilde{q})$ and $(1,0)$ which satisfies $x'+pq'=0$, and so the non-zero\footnote{If the sum of two vectors $a,b\in \mathcal{\bar{K}}$ equals zero then $a=-b\in \mathcal{\bar{K}} \cap (-\mathcal{\bar{K}})=\{0\}$.} vector $(x',q')$ belongs to $ \mathcal{\bar{K}} \cap (1,p)^{\perp}$; thus, item \ref{perp=0} implies item \ref{Lagain}.
That item \ref{Lagain} implies item \ref{E_Q(f)=p} is part of \citet*[Lemma 8]{HugonKram:04}, and that item \ref{E_Q(f)=p} implies item \ref{Pagain} follows simply from the definition of $ \mathcal{X}(x,q)$. Finally, let us prove that item \ref{Pagain} implies item \ref{Lagain}.  Let $p$ be an arbitrage-free price and assume that $(x,q)\in \mathcal{\bar{K}}$ and $x+pq\leq 0$, let $X\in \mathcal{X}(x,q)$, $X=x+H\cdot S$ and define $X':=-qp+H\cdot S$. Then $X'_T+qf$ is the sum of the two non-negative quantities $-qp-x$ and $X_T+qf$. It follows that $X'$ belongs to  $\mathcal{X}(-qp,q)$, and so by no arbitrage $X'_T+qf=0$, which implies $x+qp=0$ and $X_T+qf=0$. This proves that if $(x,q)\in \mathcal{\bar{K}}$ then $x+qp \geq 0$ (so $(1,p)\in  -\mathcal{\bar{K}}^o=\mathcal{\bar{L}})$, and that if  $x+qp=0$ and $X\in \mathcal{X}(x,q)$ then $X_T+qf=0$. Since, we can assume that such an $X$ is workable (thanks to lemma \ref{existswork}), it follows that if $x+qp=0$ then $-X\in \mathcal{X}(-x,-q)$, proving that if $(x,q)$ is in $\mathcal{\bar{K}}$ and is orthogonal to $(1,p)$ then  $(x,q)$ belongs to $\mathcal{\bar{K}} \cap (-\mathcal{\bar{K}})=\{0\}$.

Finally, let us assume that  $\mathcal{\bar{K}}^p(x)$ it non-empty, and show that $ \mathcal{\bar{K}} \cap (1,p)^{\perp}$ is its recession cone, so that \eqref{RecConeBdd} will  conclude the proof.
Since $\mathcal{\bar{K}}^p(x)$ is the set of points $(a,b) \in \mathcal{\bar{K}}$ whose dot product with $(1,p)$ equals $x$, $ \mathcal{\bar{K}} \cap (1,p)^{\perp}$ is contained in the recession cone of $\mathcal{\bar{K}}^p(x)$. For the opposite inclusion assume that $(a,b)$ belongs to $\mathcal{\bar{K}}^p(x)$ and $(c,d)$  belongs to its recession cone, then for any $n$,   $(a,b)+n(c,d) $ belongs to $\mathcal{\bar{K}}^p(x)$, so $(c,d)$ is orthogonal to $(1,p)$ and $(c,d)= \lim_n (a+nc,b+nd)/n \in  \mathcal{\bar{K}}$. $\Box$
 \newline

      \section{Upper semicontinuity of $u$}
 \label{u upper semi-continuous}
 
    Given an arbitrary vector $(y, r)\in \r^{n+1}$ , we denote by $\mathcal{Y}(y,r)$ the set of
non-negative super-martingales $Y\in \mathcal{Y}(y)$ such that the inequality
\be
\label{dualsetrend}
\mathbb{E}[Y_T (X_T + qf)] \leq xy + qr
\ee
holds true whenever $(x, q)\in \mathcal{\bar{K}}$ and $X \in \mathcal{X}(x, q)$.
Clearly this set will be empty if $(y,r)\notin \mathcal{\bar{L}}$, and it coincides with the set defined in \citet*{HugonKram:04} since asking that (\ref{dualsetrend}) holds for all $(x, q)\in \mathcal{\bar{K}}$ is equivalent to asking that it holds for all $(x, q)\in \mathcal{K}$. 
Let $\mathcal{D}(y,r)$ be the set of the positive random variables dominated by the final value of some element of  $\mathcal{Y}(y,r)$, i.e., 
 \be
 \label{D}
\mathcal{D}(y,r):=\{h\in L_+^0(P) : h\leq Y_T \textrm{ for some } Y \in \mathcal{Y}(y,r) \} ,
\ee
and define
\be
\label{C}
\mathcal{C}(x,q):=\{g\in L_+^0(P) : g\leq X_T+qf \textrm{ for some } X \in \mathcal{X}(x,q)\} . 
\ee
 Analogously to \eqref{primalSetC} and following \citet*{KramSch:99} we define
 \be
 \label{D1}
\mathcal{D}(y):=\{h\in L_+^0(P) : h\leq Y_T \textrm{ for some } Y \in \mathcal{Y}(y) \} .
\ee

 Given two sequences of functions $(g_n)_{n\geq 1}$ and $(f_n)_{n\geq 1}$, we will say that  $(f_n)_{n\geq 1}$ is a \emph{forward convex combination} of $(g_n)_{n\geq 1}$ if, for every $n$, $f_n$ is a (finite) convex combination of $(g_k)_{k\geq n}$.

\begin{lemma}  
\label{fatouconv}
Assume that conditions (\ref{na}) and (\ref{domination}) hold.
Let $g_n \in \mathcal{C}(x_n, q_n)$ for some sequence $(x_n, q_n)$ converging to $(x,q)$.
If a forward convex combination of $(g_n)_{n\geq 1}$ converges almost surely to a random variable $g$, then $g \in \mathcal{C}(x,q)$.
Analogously if a forward convex combination of  $h_n \in \mathcal{D}(y_n, r_n)$ is converging almost surely to  $h$ and $(y_n, r_n) \to (y,r)$, then $h \in \mathcal{D}(y,r)$. 
\end{lemma}
   \textsf{ PROOF. } The first statement follows applying Fatou's lemma to the converging forward convex combination of $(g_n)_{n\geq 1}$ to obtain 
\begin{align}
\label{ineqinC}
 \mathbb{E}[gh]\leq xy+qr \quad \text{ for all } (y,r)\in \mathcal{\bar{L}} \text{ and } h\in \mathcal{D}(y,r),
\end{align} 
 so that $g\in \mathcal{C}(x,q)$ follows `from \citet*[Proposition 1, item 1]{HugonKram:04}' (although stated only for $(x,q)\in \mathcal{K}$, this automatically implies that it holds for any $(x,q)\in \mathcal{\bar{K}}$: indeed if  some function $g\in L_+^0(P)$ satisfies  \eqref{ineqinC} for some $(x,q)\in \mathcal{\bar{K}}$  then 
$$ \mathbb{E}[(g+1)h]\leq xy+qr+ \mathbb{E}[h]\leq (x+1)y+qr \quad  \text{ for all } (y,r)\in \mathcal{\bar{L}} \text{ and } h\in \mathcal{D}(y,r),$$
 and thus, since\footnote{Recall that the convex cone $\mathcal{\bar{K}}$ contains $(1,0)$ in its interior (see  \citet*[Lemma 1]{HugonKram:04}).} $(x+1,q)\in \mathcal{K}$, from  \citet*[Proposition 1, item 1]{HugonKram:04}  we get  that $g+1\in \mathcal{C}(x+1,q)$, which trivially implies that  $g\in \mathcal{C}(x,q)$).

 The proof of the second statement follows analogously `from   \citet*[Proposition 1, item 2]{HugonKram:04}'; while this item is stated only for $(y,r)\in \mathcal{L}$, clearly\footnote{The proof of this item does not rely on any other lemma in \citet*{HugonKram:04}, and is short.} it is actually valid with the same  proof for any $(y,r)\in \mathcal{\bar{L}}$. $\Box$
 \newline

      \begin{remark}
      \label{Ynonempty}
          Assume that conditions (\ref{na}) and (\ref{domination}) hold. Then $\mathcal{Y}(y,r)\neq \emptyset$ if and only if $(y,r)\in \mathcal{\bar{L}}$.
      \end{remark}
          \textsf{ PROOF. } One implication is trivial. For the vice versa, observe first that $\mathcal{D}(y,r)$ is not empty if  $(y,r)\in \mathcal{L}$, as this follows from \citet*[Lemmas 8 and 9]{HugonKram:04}. If $(y,r)$ is an arbitrary point in $ \mathcal{\bar{L}}$ then take $(y_n,r_n)\in \mathcal{L}$ such that $(y_n,r_n) \to (y,r)$ and choose $h_n\in\mathcal{D}(y_n,r_n)\subseteq \mathcal{D}(\sup_n y_n)$. We can then apply Komlos' lemma to find a forward convex combination of $(h_n)_{n\geq 1}$ converging to some random variable $h$, and apply Lemma \ref{fatouconv} to show that $h\in \mathcal{D}(y,r)$. $\Box$
 \newline

 The following theorem allows us to control the behavior of the maximal expected utility function $u$ on the boundary of its domain. 
  
      \begin{theorem}
      \label{uusc}
Under the assumption of Theorem \ref{afp=mubp} the function 
 $u:\r^{n+1}\rightarrow [-\infty,\infty)$ is upper semi-continuous and for all $(x,q) \in \{u>-\infty\}$ there exists a unique  maximizer to (\ref{prrend}). 
  \end{theorem}
To prove Theorem \ref{uusc} we will need the following fact, which was proved in the second half\footnote{Starting just after formula (25).} of the proof of \citet*[Lemma 1]{KramSch:03} under the assumption that $\lim_{x\to\infty} u(x)/x=0$, which is satisfied under our hyphotheses since\footnote{This simply follows from the biconjugacy relationship between $w$ and  $\tilde{w}$, proved in \citet*[Theorem 1]{KramSch:03}, and our assumption that $ \tilde{w}(y) <\infty \textrm{  for all } y>0 .$} $u'(\infty)=0$ (as stated in \citet*[Theorem 2]{KramSch:03}) and  $\lim_{x\to\infty} u(x)/x=u'(\infty)$ (by l'Hospital's rule).

 \begin{remark}
      \label{U(C)unifInt}
Assume that conditions (\ref{na}),(\ref{inada})  hold, and that
 \bd
 \tilde{w}(y) <\infty \textrm{  for all } y>0 .
 \ed
Let  $(g^m)_{m \geq 1} \subseteq \mathcal{C}(x)$ for some $x>0$; then $(U(g^m)^+)_{m \geq 1}$ is uniformly integrable.

  \end{remark}

  \textsf{ PROOF OF THEOREM \ref{uusc}. }  To prove that $u$ is upper semi-continuous let $(x^k,q^k)$ be a sequence converging to $(x,q)$, assume without loss of generality that $u(x^k,q^k)>-\infty$ and take $X^k \in \mathcal{X}(x^k,q^k)$ such that $u(x^k,q^k)-1/k \leq \mathbb{E}[U(X_T^k+q^kf)]$. Define $g^k:=X_T^k+q^kf$ and $s:=\limsup_k u(x^k,q^k)$. Passing to a subsequence without re-labelling we can assume that $u(x^k,q^k)$ converges to $s$, and so 
  \be
  \label{ineq1}
  s= \lim_m \left( \inf_{k\geq m} u(x^k,q^k)-1/k \right) \leq \lim_m \left(\inf_{k\geq m} \mathbb{E}[U(g^k)]\right) .
  \ee
  If $x_0^{'}$ is the initial value of the process $X^{'}$ appearing in (\ref{domination}) and $\bar{x}$ is the supremum of the bounded sequence 
  \bd \textstyle
  \left(x^k+ x_0^{'} \max_{1 \leq j \leq n}|q_j^k| \right)_{k \geq 1} ,
  \ed
   then assumption (\ref{domination}) implies that, for every $m$, $g^m \in \mathcal{C}(\bar{x})$.
   
  We can then apply Komlos' lemma to find  a forward convex combination $(\tilde{g}^k)_{k\geq 1}$ of $(g^k)_{k\geq 1}$ which is converging almost surely to some random variable $g$. Then Lemma \ref{fatouconv} gives that $g\in \mathcal{C}(x,q)$, and Jensen inequality yields
  \be
  \label{ineqjen} \textstyle
  \inf_{k\geq m} \mathbb{E}[U(g^k)] \leq  \mathbb{E}[U(\tilde{g}^m)] .
  \ee
   Since $(\tilde{g}^m)_{m \geq 1} \subseteq \mathcal{C}(\bar{x})$, Remark \ref{U(C)unifInt} says that the sequence $(U(\tilde{g}^m)^+)_{m \geq 1}$ is uniformly integrable.  Thus, Fatou's lemma and inequalities (\ref{ineq1}) and (\ref{ineqjen}) imply that
  \be
  \label{ineq3} \textstyle
 \limsup_k u(x^k,q^k)=s\leq \limsup_m \mathbb{E}[U(\tilde{g}^m)] \leq \mathbb{E}[U(g)]\leq u(x,q),
  \ee
  which says that $u$ is upper semi-continuous.
   If $u(x,q)>-\infty$ then one can take $(x^k,q^k)=(x,q)$ in the above, and so (\ref{ineq3}) gives that there exists a maximizer to (\ref{prrend}). Uniqueness follows from the strict concavity of $U$.  $\Box$
 \newline

\section{Proof of the first main theorem}
\label{Proof of the main theorem.}
   
We will repeatedly need the following simple observation.
 \begin{lemma}
\label{uincr}
 Under the assumptions of Theorem \ref{afp=mubp}, if  $(\tilde{x},\tilde{q}) \in \{u>-\infty\}$ and $  (x,q) \in
   \mathcal{\bar{K}} \setminus \{0\}$ then $u(\tilde{x},\tilde{q})<u(\tilde{x}+x,\tilde{q}+q)$.
\end{lemma}
        \textsf{ PROOF }
 Theorem \ref{uusc} yields the existence of a maximizer $X(x,q)$ to (\ref{prrend}), and \citet*[Lemma 7]{HugonKram:04} provides the existence of a $\tilde{X} \in
\mathcal{X}(\tilde{x},\tilde{q})$ such that $P(\tilde{X}_T+\tilde{q}f
>0)>0$.
     Thus the following inequality holds 
     \be
     \label{ineq}
     U(X_T(x,q)+\tilde{X}_T+(q+\tilde{q})f)\geq U(X_T(x,q)+qf) ,
     \ee
    and it is a strict inequality with strictly positive
    probability. 
The thesis follows integrating the two sides of
    (\ref{ineq}), since the right hand side is
    integrable and has integral $u(x,q)$, and the left hand side has
    integral at most $u(\tilde{x}+x,\tilde{q}+q)$.
$\Box$
 \newline

   In the next proof we will use \eqref{RecConeBdd} without further mention.
\newline

        \textsf{ PROOF. OF THEOREM \ref{afp=mubp}}  
     Let $p$ be an arbitrage-free price.  Since $(1,0)\in \K$, the set $\mathcal{\bar{K}}^p(1)$ is non-empty, and Lemma \ref{xacc0} gives that it is bounded. Since \citet*[Lemma 6 ]{HugonKram:04} gives that $\mathcal{\bar{K}}^p(1)$  is closed, the function $u$, which  is  upper semi-continuous (Theorem \ref{uusc}), when restricted to the compact set $\mathcal{\bar{K}}^p(1)$  attains its maximum at some point $(x,q)$, so that $p\in \mathcal{P}(x,q)$. Since $u(x,q)\geq u(1,0)>-\infty$, it follows that
    \bd
    \qquad  \mathcal{P}\subseteq \bigcup_{(x,q)\in \{ u >-\infty \}}  \mathcal{P}(x,q;U)  .
    \ed
     Vice versa, assume that $p$ is not an arbitrage-free price and that $u(\tilde{x},\tilde{q})\in \r$, set $y:=\tilde{x}+\tilde{q}p\in \r$ so that  $(\tilde{x},\tilde{q})\in \mathcal{\bar{K}}^p(y)$, and let's prove that $(\tilde{x},\tilde{q})$ is not a maximizer of $u$ on $\mathcal{\bar{K}}^p(y)$, so that $p \notin \mathcal{P}(\tilde{x},\tilde{q})$.
     By Lemma \ref{xacc0} there exists a non-zero
     $(x,q)$ in $\mathcal{\bar{K}} \cap    (1,p)^{\perp}$, the recession cone of $\mathcal{\bar{K}}^p(y)$. 
Applying Lemma \ref{uincr} it follows that $u(\tilde{x},\tilde{q})<u(\tilde{x}+x,\tilde{q}+q)$ and so, since $(\tilde{x}+x,\tilde{q}+q)$ belongs to $ \mathcal{\bar{K}}^p(y)$, $(\tilde{x},\tilde{q})$ is not a maximizer of $u$ on $\mathcal{\bar{K}}^p(y)$.  $\Box$
 \newline

 \section{Consequences and related results}
 \label{Consequences and related results}
  As we have seen in Section \ref{counterexample}, to span all
 arbitrage-free prices  it is not  enough  in general to consider marginal  prices based at points in the interior of $ \{ u >-\infty \} $.  This warrants the study of
 $\mathcal{P}(x,q)$ in the case where $(x,q)$ belongs to the boundary
 of $ \{ u >-\infty \} $; in particular we ask which properties that
 hold in the case where $(x,q)$ belongs to the interior $\mathcal{K}$
 of $ \{ u >-\infty \} $ are still true when $(x,q)$ belongs to the
 boundary $\partial \mathcal{K}=\partial \{ u >-\infty \} $.

 First of all, if $(x,q)$ belongs to the interior of $ \{ u >-\infty
 \} $, $\mathcal{P}(x,q)$ is never empty, it is a singleton if and
 only if $u$ is differentiable at $(x,q)$, and can be `computed' using
the sub-differential $\partial u(x,q)$ (as mentioned in \citet*[Remark 1]{HugonKram:04}).

  If $(x,q)$ belongs  to the boundary of $ \{ u >-\infty \} $, $\mathcal{P}(x,q)$ can be
 empty and, if it is not, it is never a singleton.  Indeed the
 following corollary of Theorem \ref{afp=mubp} specifies for which
 points $(x,q)$ the set $\mathcal{P}(x,q)$ is empty, and allows us to
 `compute it' in the same way as if $(x,q)$ belonged to the interior
 of $ \{ u >-\infty \} $. This will allow us to prove
 Theorem \ref{Pnotclosed}, which in turn  implies that
 $\mathcal{P}(x,q)$ is never a singleton.

 \begin{corollary}
   \label{thformula}
   Under the assumptions of Theorem \ref{afp=mubp}, the set $\mathcal{P}(x,q;U)$ is non-empty if and only if
   $\partial u(x,q)$ is non-empty.  If $(y,r)\in \partial u(x,q)$ then
   $y >0$, and \be
   \label{formula}
   \mathcal{P}(x,q;U)=\left\{ \, \frac{r}{y} : (y,r)\in \partial
     u(x,q)\right\}.  \ee
 \end{corollary}
 \textsf{ PROOF. }  Given $p\in\r^n$ consider the linear function
 $A_p:\r^n \to \r\times \r^n$ given by $A_p(q'):=(-q'p,q')$ and its
 adjoint $A_p^{\star}$ given by $A_p^{\star}(y,r)=r-yp$. Now fix $x\in
 \r,q\in \r^n$, and consider the proper concave function $f_p$ given
 by $f_p(q')=u((x,q)+A_p(q'))$, so that by definition $p$ is a
 marginal price at $(x,q)$ if and only if $0$ is a maximizer of $f_p$
 , i.e., iff $0 \in \partial f_p(0)$
       
 If $(y,r)\in \partial u(x,q)$ then $y >0$, since $ f(t):=u(t,q)$ is
 strictly increasing on $[x,\infty)$ (by Lemma \ref{uincr}) and $y
 \in \partial f(0)$.  We can then define $p=r/y$ and 
 \citet*[Theorem 23.9]{Rock:70} implies that \bd 0 \in A_p^{\star}( \partial u(x,q))
 \subseteq \partial f_p(0), \ed  i.e., $p\in \mathcal{P}(x,q)$.
 
 For the opposite inclusion assume $p\in \mathcal{P}(x,q)$, so that by
 definition $(x,q)\in \mathcal{\bar{K}}$ and, by Theorem
 \ref{afp=mubp} and Lemma \ref{xacc0}, $(1,p)$ belongs to $
 \mathcal{L}$. It follows from \eqref{LK>0} that
 $y:=x+qp>0$, and so \citet*[Lemma 1]{HugonKram:04} implies that \bd
 (x,q)+A_p(-q)=(y,0)\in \mathcal{K} , \ed and now another application
 of \citet*[Theorem 23.9]{Rock:70} yields \bd A_p^{\star} ( \partial
 u(x,q) )= \partial f_p(0) \ni 0 , \ed
 i.e., $r-py=0$ for some
 $(y,r)\in \partial u(x,q)$.  $\Box$ \newline

 \textsf{ PROOF OF THEOREM \ref{Pnotclosed}} By Corollary \ref{thformula} there exists a $(y_0,r_0) \in \partial
u(x,q)$ such that $r_0/y_0=p_0$. The non-empty convex closed set $\partial u(x,q)$ is unbounded
 (see \citet*[Theorem 23.4]{Rock:70}), so \eqref{RecConeBdd} says that its recession cone contains a
 non-zero element $(y,r)$. For $t\geq 0$ define 
\be
\label{ytrt}
(y_t,r_t):= (y_0,r_0)+ t (y,r) \in \partial u(x,q) ,
\ee
 and notice that Corollary \ref{thformula} implies that
$v_t:=r_t/y_t \in \mathcal{P}(x,q)$. Since  the function $(y,r) \mapsto r/y$, defined for $y>0$, sends injectively segments into segments  (see \citet*[Section 2.3.3]{BoydVan:04}) and $v_t\to r/y=:p$ as $t \to \infty$, $\{v_t: t \geq 0\}$ equals the segment $[p_0,p)$.
We now only need to prove that $p$ is not an arbitrage-free price.
From (\ref{ytrt}) it follows that, for every $(a,b) $ in $ \{ u >-\infty \} $,
\bd
u(a,b) \leq u(x,q) + (y_t,r_t)( (a,b) - (x,q)).
\ed
Dividing times $t$ and taking the limit as $t \to \infty$ yields that
\bd
 0\leq (y,r)( (a,b) - (x,q)) .
\ed
Thus, for any $\varepsilon>0$  choosing $(a,b):=(x/2,q/2)+\varepsilon (1,0) \in \mathcal{K}\subseteq \{ u >-\infty \} $  we get that $  (y,r) (x,q) \leq 2 \varepsilon y $.
Sending $\varepsilon$ to zero we obtain $  (y,r) (x,q) \leq 0$,
and so \eqref{LK>0} implies that  $ (y,r) $ does not belong to $\mathcal{L}$.
Lemma \ref{xacc0} now concludes the proof.
 $\Box$ \newline

Theorem \ref{Pnotclosed} implies that marginal prices based at a point $(x,q)$   on the boundary of $\mathcal{\bar{K}}$, if they exist, are not
  unique.  This raises the question of how to characterize
  the uniqueness of marginal prices based at all points, which is the content of the following corollary, where the function $v$ is the convex conjugate of $u$ (see \citet*{HugonKram:04}).

  \begin{corollary}
    \label{uniq}
   Under the assumptions of Theorem \ref{afp=mubp}, the following are equivalent:
    \begin{enumerate}
    \item 
\label{1} At any $(x,q) \in \r \times \r^n$ the set $\mathcal{P}(x,q;U)$ is either empty or a singleton.
\item
\label{2} $\mathcal{P}(x,q;U)$ is a singleton at any $(x,q)\in    \mathcal{K}$, and is empty at any $(x,q) \notin \mathcal{K}$.
    \item 
\label{3}
$u$ is continuously differentiable on $\mathcal{K}$ and the
      norm of its gradient $|\nabla u(x,q)|$ converges to $\infty$ as
      $(x,q)$ approaches the boundary of $\mathcal{K}$.
    \item 
\label{4}
v is strictly convex on $\mathcal{L}$.
    \end{enumerate}

  \end{corollary}

\textsf{ PROOF }
In this proof we will use Corollary \ref{thformula} without further mention.
Theorem \ref{Pnotclosed} shows that item \ref{1} is equivalent to item \ref{2}.
Since  $\partial u(\mathcal{\bar{K}})=  \mathcal{L}$ (see \citet*[Theorem 10, item 3]{Sio12Merge}), item \ref{3} is equivalent to item \ref{4} (see \citet*[Theorem 26.3]{Rock:70}). 
Finally, item \ref{2} is equivalent to item \ref{3}, as it follows from \citet*[Theorem 26.1]{Rock:70} and the fact that $\partial u$ is single-valued iff $\mathcal{P}$ is single-valued (because $y \mapsto v(y(1,p))$ is always strictly  convex when $(1,p) \in \mathcal{L}=\partial u(\mathcal{\bar{K}}))$. $\Box$ \newline


\begin{thebibliography}{}

\bibitem[Boyd and Vandenberghe, 2004]{BoydVan:04}
Boyd, S. and Vandenberghe, L. (2004).
\newblock {\em Convex optimization}.
\newblock Cambridge university press.

\bibitem[Davis, 1995]{Da:97}
Davis, M. H.~A. (1995).
\newblock Option pricing in incomplete markets.
\newblock In {\em Mathematics of derivative securities (Cambridge, 1995)},
  volume~15 of {\em Publ. Newton Inst.}, pages 216--226, Cambridge. Cambridge
  Univ. Press.

\bibitem[Delbaen and Schachermayer, 1994]{DelbSch:94}
Delbaen, F. and Schachermayer, W. (1994).
\newblock A general version of the fundamental theorem of asset pricing.
\newblock {\em Math. Ann.}, 300(3):463--520.

\bibitem[Delbaen and Schachermayer, 1997]{DelbSch:97}
Delbaen, F. and Schachermayer, W. (1997).
\newblock The {B}anach space of workable contingent claims in arbitrage theory.
\newblock {\em Ann. Inst. H. Poincar\'e Probab. Statist.}, 33(1):113--144.

\bibitem[Foldes, 2000]{Fold:00}
Foldes, L. (2000).
\newblock Valuation and martingale properties of shadow prices: an exposition.
\newblock {\em J. Econom. Dynam. Control}, 24(11-12):1641--1701.
\newblock Computational aspects of complex securities.

\bibitem[Frittelli, 2000]{Frit:00}
Frittelli, M. (2000).
\newblock Introduction to a theory of value coherent with the no-arbitrage
  principle.
\newblock {\em Finance Stoch.}, 4(3):275--297.

\bibitem[Henderson and Hobson, 2004]{HeHo09}
Henderson, V. and Hobson, D.~G. (2004).
\newblock Utility indifference pricing-an overview.
\newblock {\em Indifference Pricing. Princeton University Press}, 4.

\bibitem[Hicks, 1956]{Hick:56}
Hicks, S. J.~R. (1956).
\newblock {\em A Revision of Demand Theory}.
\newblock Oxford University Press.

\bibitem[Hobson, 2005]{Hob05a}
Hobson, D.~G. (2005).
\newblock Bounds for the utility-indifference prices of non-traded assets in
  incomplete markets.
\newblock {\em Decisions in Economics and Finance}, 28(1):33--52.

\bibitem[Hodges and Neuberger, 1989]{HodgNeuber:89}
Hodges, S. and Neuberger, A. (1989).
\newblock Optimal replication of contingent claims under transaction costs.
\newblock {\em The Review of Futures Markets}, 8:222--239.

\bibitem[Hugonnier and Kramkov, 2004]{HugonKram:04}
Hugonnier, J. and Kramkov, D. (2004).
\newblock Optimal investment with random endowments in incomplete markets.
\newblock {\em Ann. Appl. Probab.}, 14(2):845--864.

\bibitem[Hugonnier et~al., 2005]{HugonKramSch:05}
Hugonnier, J., Kramkov, D., and Schachermayer, W. (2005).
\newblock On utility-based pricing of contingent claims in incomplete markets.
\newblock {\em Mathematical Finance}, 15(2):203--212.

\bibitem[Kallsen, 2002]{Kall:02}
Kallsen, J. (2002).
\newblock Derivative pricing based on local utility maximization.
\newblock {\em Finance Stoch.}, 6(1):115--140.

\bibitem[Karatzas and Kou, 1996]{KaratKou:96}
Karatzas, I. and Kou, S.~G. (1996).
\newblock On the pricing of contingent claims under constraints.
\newblock {\em Ann. Appl. Probab.}, 6(2):321--369.

\bibitem[Kramkov and Schachermayer, 1999]{KramSch:99}
Kramkov, D. and Schachermayer, W. (1999).
\newblock The asymptotic elasticity of utility functions and optimal investment
  in incomplete markets.
\newblock {\em Ann. Appl. Probab.}, 9(3):904--950.

\bibitem[Kramkov and Schachermayer, 2003]{KramSch:03}
Kramkov, D. and Schachermayer, W. (2003).
\newblock Necessary and sufficient conditions in the problem of optimal
  investment in incomplete markets.
\newblock {\em Ann. Appl. Probab.}, 13(4):1504--1516.

\bibitem[Kramkov and S{\^{\i}}rbu, 2006]{KramSirb:06b}
Kramkov, D. and S{\^{\i}}rbu, M. (2006).
\newblock Sensitivity analysis of utility-based prices and risk-tolerance
  wealth processes.
\newblock {\em Ann. Appl. Probab.}, 16(4):2140--2194.

\bibitem[Rockafellar, 1970]{Rock:70}
Rockafellar, T. (1970).
\newblock {\em Convex analysis}.
\newblock Princeton Mathematical Series, No. 28. Princeton University Press,
  Princeton, N.J.

\bibitem[Siorpaes, 2013]{Sio12Merge}
Siorpaes, P. (2013).
\newblock Optimal investment and price dependence in a semi-static market.
\newblock {\em arXiv:1303.0237}.

\end{thebibliography}
\bibliographystyle{apalike}

\end{document}